\documentclass[usegraphicx,usenatbib]{mn2e}                              
\usepackage{graphicx}
\usepackage{txfonts}
\usepackage{natbib}
\usepackage{epsfig}

\title{Long-term infrared variability of the UX\,Ori-type star SV\,Cep}

\author[A. Juh\'asz et al.]{A.~Juh\'asz$^{1}$\thanks{E-mail:ajuhasz@konkoly.hu}, 
T.~Prusti$^{2}$, P.~\'Abrah\'am$^{1}$,C.\,P.~Dullemond$^{3}$\\
$^{1}$Konkoly Observatory of the Hungarian Academy of Sciences,
           P.O. Box 67, H-1525 Budapest, Hungary \\
$^{2}$Space Science Department of ESA, ESTEC, Posbus 299, 2200 AG
	   Noordwijk, The Netherlands \\
$^{3}$Max-Planck-Institut f\"ur Astronomie, K\"onigstuhl 17, D-69117, 
    	Heidelberg, Germany}
\begin{document}

\date{Recieved/Accepted}

\maketitle

\label{firstpage}

\begin{abstract}
We investigate the long-term optical-infrared variability of SV\,Cep, and
explain it in the context of an existing UX Ori (UXOR) model.
A 25-month monitoring programme was completed with the Infrared Space
Observatory in the 3.3--100\,$\mu$m wavelength range. Following a careful data
reduction, the infrared light curves were correlated with the variations of
SV\,Cep in the V-band.
A remarkable correlation was found between the optical and the far-infrared
light curves. In the mid-infrared regime the amplitude of variations is lower,
with a hint for a weak anti-correlation with the optical changes. In order to interpret
the observations, we modelled the spectral energy distribution of 
SV\,Cep assuming a self-shadowed disc with a puffed-up inner rim, using a
2-dimensional radiative transfer code. We found that modifying the height of the
inner rim, the wavelength-dependence of the long-term optical-infrared variations
is well reproduced, except the mid-infrared domain. The origin of variation of the 
rim height might be fluctuation in the accretion rate in the outer disc.
In order to model the mid-infrared behaviour we tested to add an optically thin 
envelope to the system, but this model failed to explain the far-infrared variability.
Infrared variability is a powerful tool to discriminate between  
models of the circumstellar environment. The proposed mechanism of variable rim height 
may not be restricted to UXOR stars, but might be a general characteristic of
intermediate-mass young stars.
\end{abstract}


\section{Introduction} 
\label{sect:Intro}

UX\,Orionis stars are intermediate mass pre-main sequence objects defined by
the 'UXOR phenomenon': short (days to weeks) eclipse-like dimming at optical
wavelengths. Earlier explanations of the phenomenon invoked a large number  of
protocometary clouds or cometary bodies \cite{Grady}, or assumed that UX\,Ori
objects are surrounded by an almost edge-on protoplanetary disc, where hydrodynamic
fluctuations of the disc surface can cause dust filaments passing in front of
the star \cite{Bertout}. A more recent model of  Dullemond et al.
\cite{Dullemond} proposes that UX\,Ori stars harbour {\it self-shadowed} discs
(the type of  Group II in the classification scheme of Meeus et al.~2001),
where a puffed-up rim at the inner edge of the disc casts shadow over the whole
disc. Small hydrodynamical fluctuations in the puffed-up rim along the
line-of-sight may cause the  extinction events seen in UXORs.
 
The fact that UXORs may also exhibit long-term variability received much less 
attention so far in the literature. Rostopchina et al.~\cite{Rostopchina} reported
that brightness variations on a timescale  of several years with an amplitude
of 0.2--1.0\,mag can be a characteristic of the UXOR class. They conclude that a
possible reason of these brightness variations can be changes of large scale
structures in the protoplanetary discs around these stars.

It is an important test for the self-shadowed disc hypothesis  (and in general
for any UXOR model) to check whether the observed  long-term variations could
be interpreted in the framework of the model.  One possibility would be to
assume that some effects in the inner rim (e.g. weak accretion) change the rim's 
height along a significant  fraction of its perimeter  (rather than along the line-of-sight
only), thus the eclipses may last longer.  Another hypothesis is that the disc
is not completely self-shadowed but  has a moderately flared  outer part (tenuous enough
to ensure that the star is still visible), and in these outer regions dense 
filaments pass in front of the star. Due to the larger Keplerian times, these
eclipses in the outer disc last significantly longer than the normal UXOR
events.

A powerful tool to discriminate between such possibilities is looking for
variability in the thermal infrared emission of the circumstellar disc. If the
optical changes are due to line-of-sight effects then no infrared variability
is expected, since the total irradiation of the disc by the star, and
consequently the disc's total integrated infrared emission, is constant. But if
a considerable re-structuring of the inner rim occurs,  it could affect the
illumination of the disc by the star,  leading to changes in the infrared
emission, too.

In this paper we report on the results of an infrared monitoring programme of
the UX Orionis-type star SV\,Cephei. SV\,Cep is an A0-B9 star  (Rostopchina et
al.~2000) with an age of $4\cdot10^{6}$ yrs (Natta et al.~1997).  
Its distance is estimated to be 700\,pc, with some uncertainty
\cite{Kun}.  The brightness of SV\,Cep varies on a long timescale ($\sim$4000
days) between 10.5 and 11.5\,mag in the V-band, with some hints for a
quasi-periodic behaviour (Rostopchina et al.~2000). The infrared monitoring
programme was carried out with the {\it Infrared Space Observatory} (ISO) for
25 months. It is important that near-simultaneous optical data  were available
for the same period in the literature.

In Sect.~2 we summarize the ISO observations and data reduction. We present a
new algorithm for the evaluation of the ISOPHOT photometry, which was 
successfully used to reduce the relative uncertainties within the light curves.
In Sect. 3 we present the results, then in Sect. 4 we model the  light
variations assuming a self-shadowed disc with a puffed-up inner rim  and a
tenuous, moderately flared outer part, using a radiative transfer code.
Section~5 provides a summary of the work.

\section{Observations and data reduction} \label{sec:obs}
\subsection{Observations} 

SV Cep was observed with ISOPHOT \cite{lemke}, the photometer on-board the 
{\it Infrared Space Observatory}, between 1996 and 1998. The measurements were
part of a monitoring programme dedicated to  investigate the infrared
variability of UXORs. Observations were performed at 12 wavelengths  between
3.3 and 100\,$\mu$m, at 13-19 epochs depending on the wavelength.  The
filters belong to three ISOPHOT subsystems: the P1 detector (3.3--15\,$\mu$m),
the P2 detector  (20 and 25\,$\mu$m), and the P3 detector (60 and 100\,$\mu$m).
Observations were performed using the Astronomical Observing Template PHT03.
The source ('ON') and the background ('OFF') positions were measured
separately, always starting with the 'OFF' position. Measurements with
all filters  belonging to a specific detector (P1, P2 or P3) were performed first,
then a calibration measurement on the on-board Fine Calibration Source (FCS)
was obtained using the last filter. The log of the monitoring observations  is
presented in Tab.\,\ref{tab:ISOobs}.

In addition, two 3x3 mini-rasters at 150 and 200\,$\mu$m  with the ISOPHOT-C200
camera (ISO\_id: 56201201), as well as  a 2--12\,$\mu$m spectrophotometric 
measurement with the ISOPHOT-S subinstrument (ISO\_id: 56201203) were obtained
on May 31, 1997.

\subsection{Standard data processing} 
\label{stdproc}

We processed all observations using the Phot Interactive Analysis V10.0 (PIA) 
\cite{Gabriel} following the standard data reduction scheme.  First,
non-linearity correction was performed on the integration ramps, then the
two-threshold deglitching method was applied to remove cosmic particle hits.
With a first order polinomial fit to each ramp, we evaluated the signal value
and reached the Signal per Ramp Data  (SRD) level. Here reset interval
correction was performed, dark current was subtracted and  cosmic particle hits
were again checked. We took the average of the signal values to reach the next
data processing level, which is called Signal per Chopper Plateau (SCP). Since
the signal in many cases did not stabilize, we averaged only the
stable part of the measurement when possible.  The final step was the power calibration,
which could be carried out by adopting either the actual or the default 
responsivity value. The actual responsivity is derived from the FCS
measurement, while the default one is a tabulated value, computed
from data over the whole ISO mission. We decided to use default responsivity for the
P1 and P2 detectors,  and actual responsivity for filters belonging to the P3
detector. 

\begin{figure} 
\begin{center}  
\includegraphics[width=8.0cm]{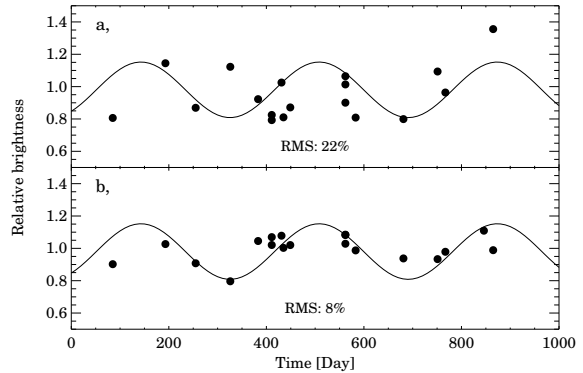}
\caption{Comparison of the standard and the new data processing schemes at
12\,$\mu$m. The solid line is a normalized curve interpolated among DIRBE measurements.
Filled dots are the normalized ISOPHOT
measurements: {\it a)} reduced with the standard method 
 and {\it b)} reduced with our new data processing scheme. The
second last ISOPHOT measurement  on the upper figure is too high, and falls out
of the figure range.} 
\label{fig:comp12} 
\end{center}  
\end{figure}

According to the ISOPHOT Handbook (Laureijs et al. 2003),  the typical absolute 
photometric
accuracy of the type of OFF/ON measurements is 40\%, 15\%,
and 15\% for the P1, P2, and P3 detectors, respectively, in the brightness
range of SV\,Cep. Due to the nearly identical instrument setup the 
relative photometric uncertainties within the light
curve, however, may be lower than these values.
In order to have an independent estimate of the relative flux
uncertainties, COBE/DIRBE data were analysed. We
compared the temporal brightness variations of the ISOPHOT background
measurements with the corresponding DIRBE data  extracted for the same
coordinates and solar elongations (see Fig. \ref{fig:comp12}{\it a}).  Though
the  DIRBE measurements are presented as surface brightness values in [MJy/sr]
and the  ISOPHOT data are in flux density units of [Jy] the ratio between the
two datasets depends only on the beam size  of ISOPHOT and should be constant.
Thus the standard deviation of this  ratio around an average value can be
adopted as the relative photometric  error of the normalized light curve. The
resulting values are 22\%, 13\%, and 10\% for the P1, P2, and  P3 detectors,
respectively.

The far-infrared mini-rasters at $150$ and $200\,\mu$m were processed 
with PIA in the standard way up to the calibrated map level (for 
calibration the FCS measurements were used). Flux extraction was performed
by fitting the measured ISOPHOT point-spread function to the brightness
distribution in the map. The ISOPHOT-S observation was reduced following 
the processing scheme described by \'Abrah\'am \& Kun~\cite{phts}, which
checks for memory effects and possible off-centre positioning of the source, 
subtracts an estimated background spectrum, and adopts realistic 
error bars. 

\subsection{A new processing scheme}
\label{newscheme}

When looking for variability, usually the highest possible photometric accuracy is needed. 
We developed an independent new processing scheme for the
ISOPHOT photometry, which takes into account that in a monitoring programme  a
uniform observing strategy  (filter sequence, heating power
of the FCS which determines its optical emission, etc.) is 
adopted at each epoch. This similarity of the observing sequences helps to
handle the detector transients in a better way,  and achieve a higher relative
accuracy.  The basic idea is the following: due to the detector transients  the
signals are usually not stabilized during the measurement time,  but because of
the identical observing strategy and illumination history,  the shapes of the
transient  curves are self-similar on the different days.  In our new data
reduction method we do not attempt to determine 'stabilized signal
level'  (e.g. by extrapolating the transient curves, or by averaging  its
stable-looking part), but measure relative flux changes from one day to the
other via direct comparison of the complete unstabilized  transient curves. 

\begin{figure}
\begin{center} 
\includegraphics[width=80mm]{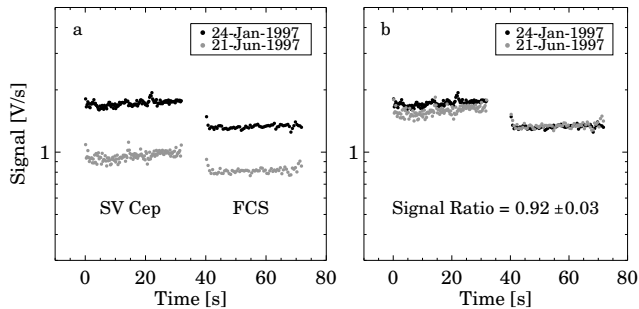}
\caption{Determining relative signal ratios in the case of two 
'ON' measurements at 100\,$\mu$m, following the principles of our new
processing scheme (Sect.\,\ref{newscheme}).
In the first step the FCS signals are scaled together, 
in order to correct for responsivity differences, then the ratio of the 
corrected SV\,Cep signal curves is computed. 
The accuracy of the relative flux difference is a few percent.}
\label{fig:scalfig}
\end{center} 
\end{figure}

As outlined in Fig.~\ref{fig:scalfig}, first we compare the FCS signals, and --
since they correspond to the same optical  power -- determine the responsivity
difference between the two epochs.  In practice,  this comparison is done by
computing a scaling factor between the -- usually unstabilized -- signal
sequences at the SRD level. This factor is applied also on the sky measurements,
since they are also affected by responsivity changes. Then the ratio of
the  corrected sky signals of the two  days is computed, which should now
reflect true  brightness variations only. This way a relative  light curve can
be built up with respect to a reference epoch.  Absolute fluxes can be obtained
by multiplying this normalized curve with the calibrated flux value of the
reference day, as determined in Sect.\,\ref{stdproc}.

There were monitoring sequences, however, when the FCS heating power was not 
kept constant for each epoch. For such cases we propose two solutions. In {\it
Method~A}, which can be applied only  for background measurements, we divide
the sequence into as many groups as many different heating power values exist
in the sequence.  Then we choose a reference day in each group and build up the
normalized light curves of these  groups separately, as described above.
Extracting COBE/DIRBE data for the sky positions and solar elongations of the
ISOPHOT background measurements, for each group  the mean ISOPHOT/DIRBE ratio
is computed. Since this ratio should be the same for all groups, we 
divide the normalized light curve of a group by the ratio between the
mean ISOPHOT/DIRBE value of the group and that of the reference group. In the
final step we merge the groups to create the final light 
curve of the whole sequence.

In {\it Method~B}, which can also be applied on source measurements, 
we take into account that the optical power of the lamp differs between the two
epochs, and scale the levels of the FCS signal curves according to this power 
difference. 
The optical powers are specified in the FCS calibration files.
After this scaling the FCS transient curves can be compared as before in order
to determine the responsivity difference. The only disadvantage of this method
is that the shapes of FCS transient curves of different heating powers
may be less similar. 

\subsection{Application of the new method on SV\,Cep}

The new processing scheme produces separate OFF and ON light curves, 
whose difference shows the temporal
variability of the object. As a reference day for the final 
normalized light curves of SV\,Cep we chose January 24, 1997, 
when most measurements showed particular stability. 
We started with the rather homogeneous 100\,$\mu$m sequence (14 epochs,
see Tab.\,\ref{tab:ISOobs}), 
where each FCS measurement was carried out with the 100\,$\mu$m filter,
and not more than 3 different FCS heating powers were used. We applied 
{\it Method A} on the background measurements, including a relatively 
large correction on the first day (February 10, 1996) when the
ISOPHOT/DIRBE ratio was significantly  higher than its
average value (by more than 3$\sigma$), thus we scaled down the ISOPHOT OFF 
signal by this ratio.
Since at 100\,$\mu$m the ON and OFF signal levels were comparable and
the same FCS heating power was used for both 
source and background measurements, we decided to apply the 
correction factors derived for background measurements on the source 
measurements, too. 
At 60\,$\mu$m the observational sequence was less homogeneous, since 
the FCS calibration was carried out either with the 60\,$\mu$m or
the 100\,$\mu$m filter, and different heating powers were used. 
Thus we had to apply 
{\it Method B} in this case, and again a noticeable correction
was necessary at the first epoch.
At 20 and 25\,$\mu$m (P2 detector), and at shorter
wavelengths (P1 detector) we always used {\it Method B}.

The relative photometric uncertainties were again evaluated via comparison with 
DIRBE background measurements (see Fig.\,\ref{fig:comp12}$b$ for the 12\,$\mu$m
filter). The figure shows that our new  data processing scheme produces 
significantly lower relative photometric errors compared to the 
standard method. The errors of the final ON-OFF differential 
photometry are typically 14--15\,\% at 60 and 
100\,$\mu$m; 14--15\,\% at 20 and 25\,$\mu$m; and are about 8-9\,\% for the P1 filters
(the latter was computed at 12$\mu$m and adopted for the other P1 filters, too).

\begin{table*}
\begin{center}
\begin{tabular}{ccc@{}cccc}
\hline
 Date        & ISO\_id       & $\lambda$  		               & Aperture               &  FCS heat. pow. \\ 
	     & OFF / ON	       &  [$\mu$m]                   	      &	[$\arcsec$]	        &  OFF / ON [mW] \\ \hline
10-Feb-1996  & 08502135 / 36     & 3.3,3.6,4.8,7.3,10.0,12.0,12.8,15.0  & 	18  	       &  7.37485 / 4.55433 \\
             & 08502135 / 36     & 20,25	         		    &  	52  	       &  1.50183 / 1.86813 \\
             & 08502135 / 36     & 60,100	         		    &  	120  	       &  0.58608 / 0.68376 \vspace{-1mm}\\ 
\hline
28-May-1996  & 19302325 / 26     & 3.3,3.6,4.8,7.3,10.0,12.0,12.8,15.0  & 	18  	       &  7.98535 / 5.88523 \\
             & 19302325 / 26     & 20,25	         	   	    &  	52  	       &  1.50183 / 2.24664 \\
             & 19302325 / 26     & 60,100	         		    &  	120  	       &  0.41514 / 0.41514\vspace{-1mm} \\ \hline
28-Jul-1996  & 25500127 / 28     & 3.3,3.6,4.8,7.3,10.0,12.0,12.8,15.0  & 	18  	       &  8.00977 / 5.88523 \\
             & 25500127 / 28     & 20,25	         	      &  	52  	       &  1.50183 / 2.24664 \\
             & 25500127 / 28     & 60,100	         		    &  	120  	       &  0.41514 / 0.41514\vspace{-1mm} \\ \hline
08-Oct-1996  & 32601431 / 32     & 3.3,3.6,4.8,7.3,10.0,12.0,12.8,15.0  & 	18  	       &  8.00977 / 5.88523 \\
             & 32601431 / 32     & 20,25	         	      &  	52  	       &  1.50183 / 2.24664 \\
             & 32601431 / 32     & 60,100	         		    &  	120  	       &  0.41514 / 0.41514\vspace{-1mm} \\ \hline
03-Dec-1996  & 38300901 / 02     & 3.6,12.0			      & 	18  	       &  8.00977 / 5.88523 \\
             & 38300901 / 02     & 20,25	         	      &  	52  	       &  1.40415 / 1.79487 \\
             & 38300901 / 02     & 60	         		      &  	120  	       &  0.59829 / 0.70818\vspace{-1mm} \\ \hline
31-Dec-1996  & 41100637 / 38     & 3.6,12.0			      & 	18  	       &  8.00977 / 5.88523 \\
             & 41100637 / 38     & 20,25	         	      &  	52  	       &  1.40415 / 1.79487 \\
             & 41100637 / 38     & 60	         		      &  	120  	       &  0.59829 / 0.70818\vspace{-1mm} \\ \hline
31-Dec-1996  & 41100733 / 34     & 3.3,3.6,4.8,7.3,10.0,12.0,12.8,15.0  & 	18  	       &  8.00977 / 5.88523 \\
             & 41100733 / 34     & 20,25	         	      &  	52  	       &  1.40415 / 2.21001 \\
             & 41100733 / 34     & 60,100	         		    &  	120  	       &  0.35409 / 0.35409\vspace{-1mm} \\ \hline
21-Jan-1997  & 43101241 / 42     & 3.6,12.0			      & 	18  	       &  8.00977 / 5.88523 \\
             & 43101241 / 42     & 20,25	         	      &  	52  	       &  1.40415 / 1.79487 \\
             & 43101241 / 42     & 60	         		      &  	120  	       &  0.59829 / 0.70818\vspace{-1mm} \\ \hline
24-Jan-1997  & 43501229 / 30     & 3.3,3.6,4.8,7.3,10.0,12.0,12.8,15.0  & 	18  	       &  8.00977 / 5.88523 \\
             & 43501229 / 30     & 20,25	         	      &  	52  	       &  1.40415 / 2.21001 \\
             & 43501229 / 30     & 60,100	         		    &  	120  	       &  0.35409 / 0.35409\vspace{-1mm} \\ \hline
07-Feb-1997  & 44901743 / 44     & 3.3,3.6,4.8,7.3,10.0,12.0,12.8,15.0  & 	18  	       &  8.00977 / 5.88523 \\
             & 44901743 / 44     & 20,25	         	      &  	52  	       &  1.40415 / 2.21001 \\
             & 44901743 / 44     & 60,100	         		    &  	120  	       &  0.35409 / 0.35409\vspace{-1mm} \\ \hline
31-May-1997  & 56200647 / 48     & 3.3,3.6,4.8,7.3,10.0,12.0,12.8,15.0  & 	18  	       &  8.00977 / 5.90965 \\
             & 56200647 / 48     & 20,25	         	      &  	52  	       &  1.40415 / 2.21001 \\
             & 56200647 / 48     & 60,100	         		    &  	120  	       &  0.35409 / 0.35409\vspace{-1mm} \\ \hline
31-May-1997  & 56200845 / 46     & 3.6,12.0			      & 	18  	       &  8.00977 / 5.90965 \\
             & 56200845 / 46     & 20,25	         	      &  	52  	       &  1.40415 / 1.79487 \\
             & 56200845 / 46     & 60	         		      &  	120  	       &  0.59829 / 0.70818\vspace{-1mm} \\ \hline
31-May-1997  & 56201204 / 05     & 3.3,3.6,4.8,7.3,10.0,12.0,12.8,15.0  & 	18  	       &  8.00977 / 5.90965 \\
             & 56201204 / 05     & 20,25	         	      &  	52  	       &  1.40415 / 2.21001 \\
             & 56201204 / 05     & 60,100	         		    &  	120  	       &  0.35409 / 0.35409\vspace{-1mm} \\ \hline
21-Jun-1997  & 58302203 / 04     & 3.3,3.6,4.8,7.3,10.0,12.0,12.8,15.0  & 	18  	       &  8.00977 / 5.90965 \\
             & 58302203 / 04     & 20,25	         	      &  	52  	       &  1.40415 / 2.21001 \\
             & 58302203 / 04     & 60,100	         		    &  	120  	       &  0.35409 / 0.35409\vspace{-1mm} \\ \hline
26-Sep-1997  & 68100205 / 06     & 3.3,3.6,4.8,7.3,10.0,12.0,12.8,15.0  & 	18  	       &  8.00977 / 5.90965 \\
             & 68100205 / 06     & 20,25	         	      &  	52  	       &  1.40415 / 2.21001 \\
             & 68100205 / 06     & 60,100	         		    &  	120  	       &  0.35409 / 0.35409\vspace{-1mm} \\ \hline
05-Dec-1997  & 75100107 / 08     & 3.3,3.6,4.8,7.3,10.0,12.0,12.8,15.0  & 	18  	       &  8.00977 / 5.90965 \\
             & 75100107 / 08     & 20,25	         	      &  	52  	       &  1.40415 / 2.21001 \\
             & 75100107 / 08     & 60,100	         		    &  	120  	       &  0.35409 / 0.35409\vspace{-1mm} \\ \hline
21-Dec-1997  & 76700951 / 52     & 3.3,3.6,4.8,7.3,10.0,12.0,12.8,15.0  & 	18  	       &  8.00977 / 5.90965 \\
             & 76700951 / 52     & 20,25	         	      &  	52  	       &  1.40415 / 2.21001 \\
             & 76700951 / 52     & 60,100	         		    &  	120  	       &  0.35409 / 0.35409\vspace{-1mm} \\ \hline
11-Mar-1998  & 84604849 / 50     & 3.6,12.0			      & 	18  	       &  8.00977 / 5.90965 \\
             & 84604849 / 50     & 20,25	         	      &  	52  	       &  1.40415 / 1.79487 \\
             & 84604849 / 50     & 60	         		      &  	120  	       &  0.59829 / 0.70818\vspace{-1mm} \\ \hline
29-Mar-1998  & 86500853 / 54     & 3.3,3.6,4.8,7.3,10.0,12.0,12.8,15.0  & 	18  	       &  8.00977 / 5.90965 \\
             & 86500853 / 54     & 20,25	         	      &  	52  	       &  1.40415 / 2.21001 \\
             & 86500853 / 54     & 60,100	         		  &  	120  	       &  0.35409 / 0.35409\vspace{-1mm} \\ \hline
\end{tabular}
\caption{Log of ISOPHOT observations.}
\label{tab:ISOobs}
\end{center}
\end{table*}

\begin{table*}
\begin{center}
\begin{tabular}{lcccccc}
\hline
Date  & 3.3\,$\mu$m &3.6\,$\mu$m &4.8\,$\mu$m &7.3\,$\mu$m &10.0\,$\mu$m &12.0\,$\mu$m \\
\hline
10-Feb-1996  & 0.524$\pm$0.036 & 0.458$\pm$0.092 & 0.470$\pm$0.094 & 0.711$\pm$0.142 & 4.697$\pm$0.939 & 3.704$\pm$0.741 \\  
28-May-1996  & 0.900$\pm$0.062 & 0.645$\pm$0.129 & 0.562$\pm$0.112 & 0.819$\pm$0.164 & 5.864$\pm$1.173 & 4.621$\pm$0.924 \\  
28-Jul-1996  & 0.711$\pm$0.049 & 0.664$\pm$0.133 & 0.627$\pm$0.125 & 0.855$\pm$0.171 & 5.801$\pm$1.160 & 4.579$\pm$0.916 \\  
8-Okt-1996   & 0.631$\pm$0.044 & 0.491$\pm$0.098 & 0.506$\pm$0.101 & 0.788$\pm$0.158 & 5.954$\pm$1.191 & 5.059$\pm$1.012 \\  
3-Dec-1996   &  -  & 0.493$\pm$0.099 &  -  &  -  &  -  & 5.147$\pm$1.029 \\  
31-Dec-1996  & 0.758$\pm$0.052 & 0.643$\pm$0.129 & 0.663$\pm$0.133 & 1.009$\pm$0.202 & 6.458$\pm$1.292 & 4.713$\pm$0.943 \\  
31-Dec-1996  &  -  & 0.553$\pm$0.111 &  -  &  -  &  -  & 5.130$\pm$1.026 \\  
21-Jan-1997  &  -  & 0.546$\pm$0.109 &  -  &  -  &  -  & 5.147$\pm$1.029 \\  
24-Jan-1997  & 0.659$\pm$0.046 & 0.610$\pm$0.122 & 0.633$\pm$0.127 & 1.000$\pm$0.200 & 7.068$\pm$1.414 & 5.144$\pm$1.029 \\  
7-Feb-1997   & 0.714$\pm$0.049 & 0.617$\pm$0.123 & 0.659$\pm$0.132 & 0.959$\pm$0.192 & 7.170$\pm$1.434 & 5.162$\pm$1.032 \\  
31-May-1997  &  -  & 0.671$\pm$0.134 &  -  &  -  &  -  & 4.319$\pm$0.864 \\  
31-May-1997  &  -  & 0.596$\pm$0.119 &  -  &  -  &  -  & 4.842$\pm$0.968 \\  
31-May-1997  & 0.774$\pm$0.054 & 0.648$\pm$0.130 & 0.614$\pm$0.123 & 0.902$\pm$0.180 & 6.491$\pm$1.298 & 4.987$\pm$0.997 \\  
21-Jun-1997  & 0.752$\pm$0.052 & 0.656$\pm$0.131 & 0.635$\pm$0.127 & 0.915$\pm$0.183 & 6.558$\pm$1.312 & 4.974$\pm$0.995 \\  
26-Sep-1997  & 0.653$\pm$0.045 & 0.631$\pm$0.126 & 0.638$\pm$0.128 & 0.871$\pm$0.174 & 6.053$\pm$1.211 & 4.826$\pm$0.965 \\  
5-Dec-1997   & 0.648$\pm$0.045 & 0.553$\pm$0.111 & 0.544$\pm$0.109 & 0.932$\pm$0.186 & 5.504$\pm$1.101 & 4.629$\pm$0.926 \\  
21-Dec-1997  & 0.660$\pm$0.046 & 0.556$\pm$0.111 & 0.583$\pm$0.117 & 0.868$\pm$0.174 & 6.575$\pm$1.315 & 5.078$\pm$1.016 \\  
11-Mar-1998  &  -  & 0.543$\pm$0.109 &  -  &  -  &  -  & 4.453$\pm$0.891 \\  
29-Mar-1998  & 0.562$\pm$0.039 & 0.525$\pm$0.105 & 0.484$\pm$0.097 & 0.757$\pm$0.151 & 5.758$\pm$1.152 & 4.930$\pm$0.986 \\ 
\hline
\vspace*{1mm}\\
\hline
Date & 12.8\,$\mu$m &15\,$\mu$m &20\,$\mu$m &25\,$\mu$m &60$\mu$m & 100\,$\mu$m\\
\hline
10-Feb-1996  & 3.078$\pm$0.616 & 1.795$\pm$0.359 & 6.961$\pm$0.950 & 4.067$\pm$0.790 & 3.047$\pm$0.876 & 1.853$\pm$0.532 \\ 
28-May-1996  & 3.805$\pm$0.761 & 2.149$\pm$0.430 & 7.570$\pm$1.102 & 3.846$\pm$0.658 & 1.941$\pm$0.471 & 1.541$\pm$0.374 \\ 
28-Jul-1996  & 4.227$\pm$0.845 & 2.544$\pm$0.509 & 7.815$\pm$1.140 & 4.270$\pm$0.743 & 1.856$\pm$0.404 & 1.391$\pm$0.302 \\ 
8-Okt-1996   & 4.030$\pm$0.806 & 1.907$\pm$0.381 & 7.699$\pm$1.102 & 3.972$\pm$0.641 & 1.701$\pm$0.375 & 1.316$\pm$0.290 \\ 
3-Dec-1996   &  -  &  -  &  -  & 4.138$\pm$0.684 & 1.954$\pm$0.456 &  -  \\ 
31-Dec-1996  & 4.311$\pm$0.862 & 2.648$\pm$0.530 & 7.471$\pm$1.068 & 3.947$\pm$0.639 & 2.116$\pm$0.521 & 1.091$\pm$0.255 \\ 
31-Dec-1996  &  -  &  -  &  -  & 4.229$\pm$0.714 & 1.804$\pm$0.429 &  -  \\ 
21-Jan-1997  &  -  &  -  &  -  & 3.789$\pm$0.609 & 1.658$\pm$0.402 &  -  \\ 
24-Jan-1997  & 4.654$\pm$0.931 & 2.587$\pm$0.517 & 7.463$\pm$1.072 & 3.886$\pm$0.648 & 2.091$\pm$0.495 & 1.417$\pm$0.350 \\ 
7-Feb-1997   & 4.694$\pm$0.939 & 2.597$\pm$0.519 & 7.582$\pm$1.110 & 4.247$\pm$0.731 & 1.799$\pm$0.406 & 1.496$\pm$0.356 \\ 
31-May-1997  &  -  &  -  &  -  & 4.344$\pm$0.738 & 1.846$\pm$0.404 & 0.797$\pm$0.193 \\ 
31-May-1997  &  -  &  -  &  -  & 3.894$\pm$0.648 & 1.740$\pm$0.381 &  -   \\
31-May-1997  & 4.406$\pm$0.881 & 2.476$\pm$0.495 & 7.535$\pm$1.092 & 4.157$\pm$0.700 & 1.667$\pm$0.350 & 1.130$\pm$0.267  \\
21-Jun-1997  & 4.498$\pm$0.900 & 2.466$\pm$0.493 & 7.384$\pm$1.042 & 4.108$\pm$0.688 & 1.543$\pm$0.325 & 1.074$\pm$0.253  \\
26-Sep-1997  & 4.266$\pm$0.853 & 2.575$\pm$0.515 & 7.451$\pm$1.040 & 4.136$\pm$0.719 & 1.767$\pm$0.366 & 0.976$\pm$0.221  \\
5-Dec-1997   & 3.950$\pm$0.790 & 2.494$\pm$0.499 & 7.124$\pm$0.970 & 3.561$\pm$0.560 & 1.607$\pm$0.348 & 1.211$\pm$0.255  \\
21-Dec-1997  & 4.320$\pm$0.864 & 2.503$\pm$0.501 & 6.782$\pm$0.890 & 3.007$\pm$0.449 & 1.685$\pm$0.371 & 0.968$\pm$0.204  \\
11-Mar-1998  &  -  &  -  &  -  & 3.806$\pm$0.630 & 1.325$\pm$0.263 &  -   \\
29-Mar-1998  & 3.986$\pm$0.797 & 1.974$\pm$0.395 & 7.563$\pm$1.118 & 4.005$\pm$0.682 & 1.644$\pm$0.342 & 1.241$\pm$0.263 \\
\hline
\end{tabular}
\caption{Colour corrected flux densities of SV\,Cep in Jy. The quoted errors are the relative 
uncertainties 
of our new data processing scheme. The absolute flux errors from the 
standard method are higher by a factor of
2.5 for the P1 filters, 1.3 for the P2 filters and 2.5 for the P3 filters.}
\label{tab:flux2}
\end{center}
\end{table*}


\section{Results}

The resulting flux density values for all 19 epochs are presented in
Tab.\,\ref{tab:flux2}. The far-infrared flux densities on 
May 31, 1997 (not included in the table) were the following: 
0.664$\pm$0.186\,Jy at 150\,$\mu$m, and 0.498$\pm$0.075\,Jy at 200\,$\mu$m.
All flux values are colour corrected. Colour correction 
factors were
determined by convolving the ISOPHOT filter profiles with the shape of the
spectral energy distribution (SED) in an iterative way. 
Since the general shape of
the SED did not change considerably over the monitoring period, 
we computed correction factors for one particular
day (May 31, 1997, see below) and used these values for all other epochs.      
The quoted uncertainties in Tab.\,\ref{tab:flux2} correspond to the relative
photometric errors derived from the new processing scheme; the absolute
photometric uncertainties are the higher values given in Sect.\,\ref{stdproc}.

\subsection{Spectral energy distribution}

\begin{figure}
\centering
\includegraphics[width=88.1mm]{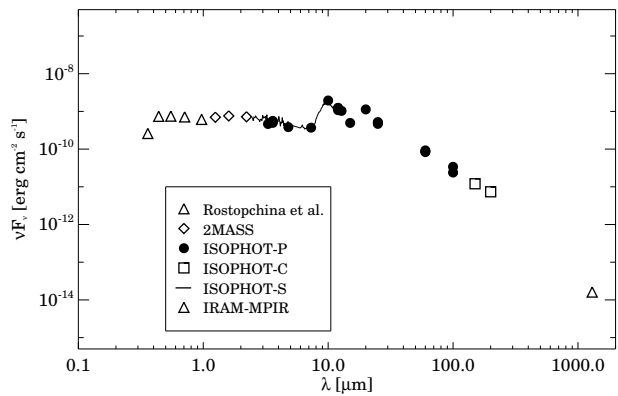}
\caption{SED of SV Cep. Flux uncertainties are usually smaller than the symbol
size. No correction for interstellar reddening was performed.}
\label{fig:sed}
\end{figure}

Before investigating the infrared variability, we constructed and studied the
SED of SV Cep. Since the star is variable, we chose a single epoch,
May 31, 1997, when a large
number of ISOPHOT measurements were obtained (including far-infrared
photometry at 150 and 200\,$\mu$m as well as mid-infrared spectrophotometry),
and constructed an SED for this reference epoch.  Figure\,\ref{fig:sed} shows
the resulting SED.   The 3.6--100\,$\mu$m data were supplemented by optical
UBVRI data, taken from Rostopchina et al.~\cite{Rostopchina} 
and interpolated for the reference
epoch, as well as by a ground based millimeter-wave  measurement
\cite{Natta:discmass}. The zero magnitude fluxes of the optical bands are taken from
Bessel et al. (1979).

At wavelengths shorter than 1\,$\mu$m the SED is dominated by the stellar
component (Fig.\,\ref{fig:sed}). Between 2 and 8\,$\mu$m a near-infrared  bump
can be seen, which is typical for many Herbig Ae stars \cite{Dullemond:hole}.
Around  10\,$\mu$m a strong silicate emission feature is present and probably
the same silicate grains are responsible for the local peak at
20\,$\mu$m.  The continuum of the SED is approximately flat (in ${\nu}F_{\nu}$)
below 25\,$\mu$m,
and decreases towards longer wavelengths approximately as 
F$_{\nu}\propto\lambda^{-1.1}$. The fact that this slope is shallower
than a Rayleigh-Jeans tail suggests the existence of cold, optically thin
components  at the outer part of the disc. In the submillimeter-wave region the
shallower slope of the SED can be the consequence of presence of big grains settled 
to the disk midplane which radiate at these wavelengths. 

\subsection{Optical-infrared light curves}

\begin{figure}
\centering
\includegraphics[width=88.1mm]{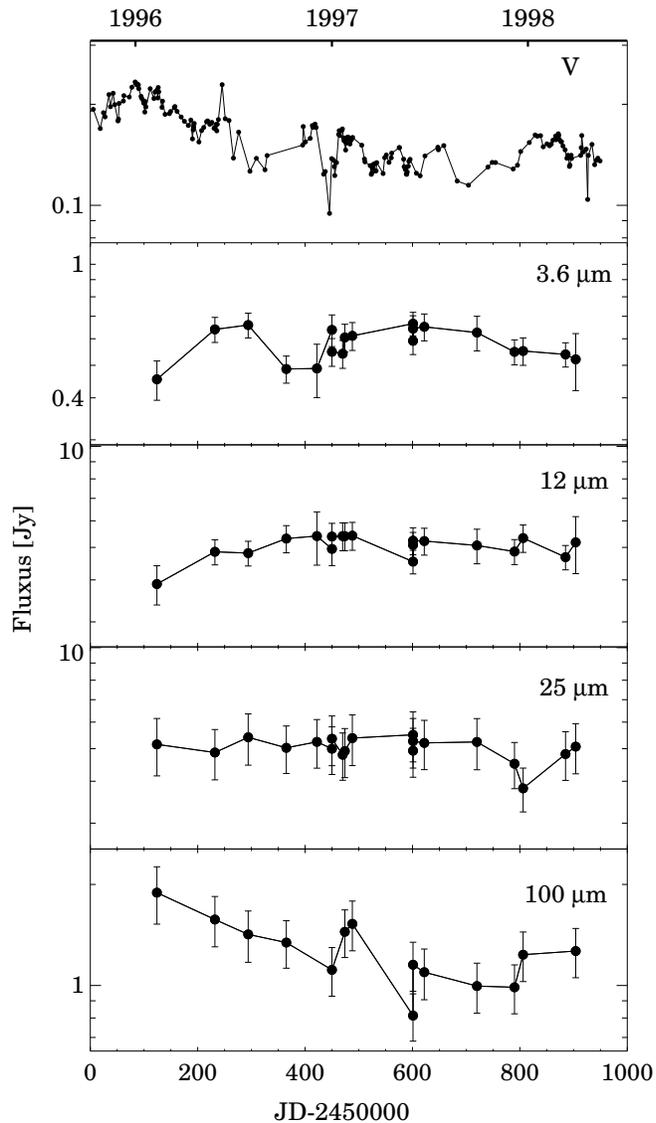}
\caption{Light curves of SV\,Cep at four representative infrared 
wavelengths, as well as in the V-band. 
Infrared flux values are taken from Tab.\,2; the V-band magnitudes are from 
Rostopchina et al. (2001). The plotted error bars correspond to the relative 
photometric uncertainties.}
\label{fig:curves}
\end{figure}

In Fig.\,\ref{fig:curves} we plotted the light curves of SV Cep at four
representative infrared  wavelengths, as well as in the V-band.  Infrared flux
values are taken from Tab.\,2; the V-band magnitudes are from  Rostopchina et
al. (2000).

In the V-band clear variability can be seen. It is not surprising, that the
brightness of SV Cep changes on a time-scale of days to weeks, which is typical
for UXORs, but the brightness varies also on a time-scale of  hundred days. The
peak-to-peak amplitude of the long-term flux variation of SV Cep is about
60\%.  At the beginning of the monitoring period, in 1996, SV\,Cep was in a
bright state, then its  brightness decreased steadily within the next 250 days.
At the beginning of 1997 there was a local maximum, followed by a shallow
decline in brightness.  At the end of 1997 the direction of the brightness
variation changed and  SV\,Cep began to brighten again.

We can also recognize variability, exceeding the formal flux
uncertainties, at 3.6\,$\mu$m.  The peak-to-peak flux variation is somewhat
smaller than in the V-band. The flux increased during the first 200 days, then,
except for a local minimum at the end of 1996, the brightness remained
constant, within the errors,  until the end of 1997. Then the flux slowly
decreased until the last measured point. At 12\,$\mu$m the peak-to-peak
variability is even smaller than at 3.6\,$\mu$m. The brightness slowly incresed
during the first 450 days, then decreased with a similar pace.  The light
curves at the other six wavelengths from 3.3 to 15\,$\mu$m (not shown)  behave
similar to the 3.6 and 12\,$\mu$m results, but the amplitude of the variability
decreases from 3.3 to 15\,$\mu$m. The flux of SV\,Cep was almost constant at 20
and 25\,$\mu$m over the observed 25 months, compared to the measurement
uncertainties.

At 100\,$\mu$m clear variability can be recognized, with a peak-to-peak change 
of about a factor of 2. The brightness decreased during the first 800 days, except for  a
local maximum at the beginning of 1997. From the beginning of 1998 the flux at
100\,$\mu$m began to increase slowly. The shape of the light curve at
60\,$\mu$m is very similar to that at 100\,$\mu$m.

\subsection{Correlations}

\begin{figure}
\centering
\includegraphics[width=88.1mm]{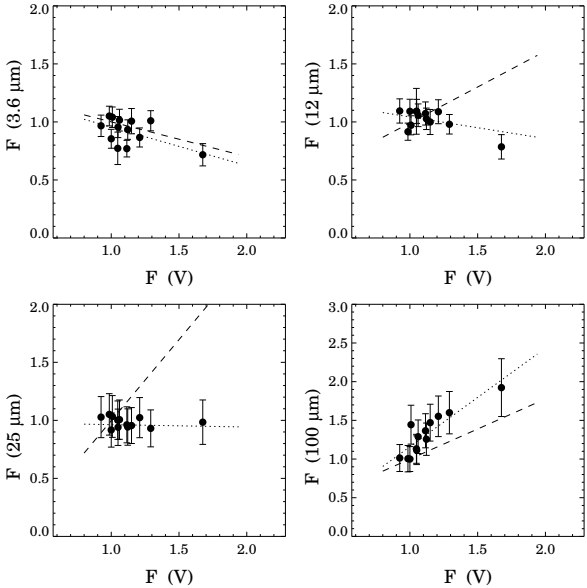}
\caption{Correlation diagrams between the optical and 
infrared light variations. All light curves are normalized
for the reference day of May 31, 1997. Dotted lines are linear fits to the 
observations; dashed lines show the same for synthetic light curves 
obtained from modelling (Sect.\,\ref{modelling}).} 
\label{fig:mod_corr}
\end{figure}

In order to investigate the relationships between the optical and infrared data
sets,  we interpolated the V-band light curve for the epochs of the ISOPHOT 
observations. Then infrared data from Tab. 2 were plotted against the
interpolated  V-band fluxes (Fig.\,\ref{fig:mod_corr}) at the same four
wavelengths as in Fig.\,\ref{fig:curves}. 

At 100\,$\mu$m a clear correlation can be seen between the optical and the
infrared flux values. The Pearson's correlation coefficient is 0.88, which
indicates   a significant linear trend in the data distribution. A similar
behaviour can be observed at 60\,$\mu$m (not plotted). At 25\,$\mu$m, and also at
20\,$\mu$m (not shown),  the correlation  with the V-band disappears, due to the
fact, that at these wavelengths SV\,Cep  does not exhibit any significant
variations (Fig.\,\ref{fig:curves}). The 3.6\,$\mu$m and 12\,$\mu$m correlation 
plots might indicate a weak anti-correlation with the V-band fluxes (Pearson's
correlation coefficients are -0.55 and -0.40, respectively). The significance of
this  anti-correlation mainly relies on the reliability of the photometric points
at the very first epoch, whose calibration was not straightforward (see
Sect.\,2.4). However, simultaneous  observations with the SWS instrument 
(the short-wavelength spectrometer on-board ISO) at 2.63 and 3.6$\mu$m 
seem to support the observed anti-correlation at mid-infrared wavelengths.

\section{Discussion}
\label{modelling}

\subsection{Variability: a new tool to explore circumstellar structure}

A number of numerical codes are available, which can model the SED of a young
stellar object by assuming a certain circumstellar geometry (e.g. flat/flared
disc, envelope) and performing radiative transfer calculations (e.g. 
Dullemond et al. \cite{Dullemond06} and references therein). In many cases, 
however, the 
model results are not unique: the SED can be fitted by assuming different 
circumstellar structures. One possibility to resolve this ambiguity is
high spatial resolution imaging. Another way -- which was first proposed by
Chiang \& Goldreich \cite{cg97} and which will be explored in this section 
-- is to make
predictions for possible flux variations in the different competing models
and confront the synthetic light curves with the observations. In practice
it is a three-step procedure:\\

\noindent{\bf Modelling the SED at one selected epoch.} 
The abovementioned stationary models can be used to fit the SED at a 
selected epoch, in order to take a snapshot of the circumstellar structure.
The best fitting model specifies, for any given wavelength, (i) 
which component of the circumstellar structure (e.g. puffed-up   inner rim, disc
atmosphere, midplane)  dominates the infrared  radiation in that part of
the spectrum;  and (ii) whether the emitting region is optically thin or 
optically thick for incident starlight. This information will be essential 
to model variability. \\

\noindent{\bf Tuning model setup parameters.}
The infrared emission of Young Stellar Objects (YSOs) can vary in time when the illumination of the
circumstellar matter by the central source is not constant. 
Disc illumination can change if (1)
the luminosity of the star varies (e.g. time-variable accretion
    onto the star; examples are the eruptive FU Ori- or EX Lup-type 
    young stars); or
(2) the circumstellar geometry is re-structured, modifying 
the illumination pattern on
the disc  (e.g. the scale height of the puffed-up inner rim -- which casts a
shadow on the outer disc -- is not stable).  
These changes can be modelled by tuning the corresponding stellar/geometrical
parameters in the setup of the stationary model of the circumstellar
environment (i.e. luminosity of the central star or scale heights of the disc). The
resulting SED can be compared with the one of the selected epoch, 
in order to predict the observable consequences of a changing stellar 
luminosity or circumstellar structure. \\

\noindent{\bf Timescales.}
The infrared emission does not react immediately to a change
of illumination. The timescale depends on the radial distance of the emitting
region from the centre and on its optical depth for the incident light.
Optically thin disc surface layer or optically thick disc interior react to a
change of illumination on completely different timescales, as computed by
Chiang \& Goldreich \cite{cg97}. 
New time-dependent codes would be necessary to compute the timescales 
related to the changes of illumination for
each wavelength correctly. However, as a first attempt, one can often 
assume that emission from an optically thin region (e.g. puffed-up rim, disc
atmosphere) reacts instantaneously to the change of illumination, while the
emission of optically thick components is practically constant on the
timescales of the monitoring programme ($\sim$years). 
With these assumptions one can predict how the infrared emission
would react, as a function of time, on variations of the illumination of the
disc. The predictions can be compared with the measured infrared light curves.
Significant discrepancies may reveal that the model assumptions on the
circumstellar structure have to be modified in order to reach a closer
match to reality.

\subsection{Modelling of SV\,Cep}

\subsubsection{Disc model}
\label{sec:disc_model}

\begin{figure}
\centering
\includegraphics[width=88.1mm]{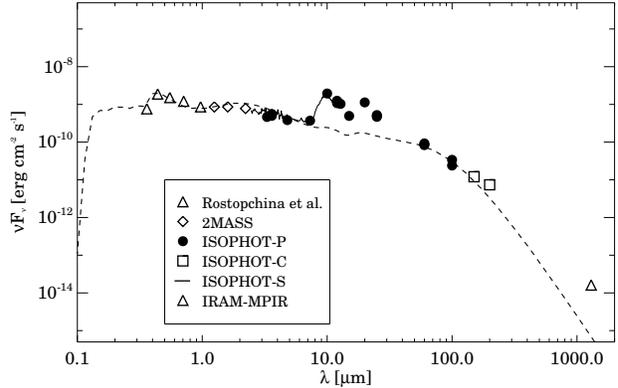}
\caption{SED of SV Cep. The measured data points are corrected for an
interstellar reddening of 0.74$^m$. Overplotted is the best fitted disc model.}
\label{fig:modsed}
\end{figure}

\noindent{\bf Modelling the SED.}
\label{sec:mod_sed_disc}
In Sect.\,\ref{sect:Intro} we asked the question whether the observed long-term
variability of SV\,Cep could be interpreted in the framework of the self-shadowed 
disc model proposed for UX\,Orionis stars by Dullemond et al. \cite{Dullemond}. 
Two possible scenarios were outlined: 
temporal changes in the inner rim's height along a significant fraction of its perimeter; 
and eclipses in a moderately flared  outer part of the disc. The latter possibility is
a line-of-sight effect which does not change the total illumination of the disc,
thus the observed infrared variability clearly excludes this option.
In the following we perform a quantitative analysis 
of the first scenario.

As a first step, we investigate whether the SED of SV\,Cep of a selected epoch
could be reproduced with a self-shadowed disc geometry. SV\,Cep is a typical
UX\,Orionis star, whose strong near-IR emission and low far-infrared fluxes are
qualitatively consistent with such a model. For the quantitative analysis we
chose the optical-to-submillimeter SED of 31 May, 1997 (plotted in 
Fig\,\ref{fig:sed}).
The model disc includes a puffed-up inner rim which  casts shadow over the 
outer  regions. The outer disc is not fixed  {\it a priori} to a flat geometry,
but may have a flaring part whose curvature is determined by the shape of the SED.
The whole disc is thought to be passive,   i.e. its energy source is absorbed
stellar radiation with no additional  release of accretion energy.

For the modelling we used the 2-D radiative transfer codes
RADICAL\footnote{http://www.mpia.de/homes/dullemon/radtrans/radical/} and
RADMC\footnote{http://www.mpia.de/homes/dullemon/radtrans/radmc/}. These
codes assume axial symmetry for the circumstellar environment and adopt
realistic opacities  to perform the radiative transfer in order to find the
dust temperature as a function of radius  and polar angle. During the modelling
procedure we first used RADMC to determine the temperature distribution in the
circumstellar disc, then we used the raytracer of RADICAL in order to produce the SED. 

For the star we assumed M$_*=3.1$\,M$_\odot$, R$_*=3.6$\,R$_\odot$,  and 
T$_*$=10200\,K (Rostopchina et al. 2000). Dust population in the disc 
was a mixture  of silicate grains with two different diameters  (0.1\,$\mu$m 
and 1.0\,$\mu$m). The size distribution was assumed to be a power law with a 
power index of $-3.5$. The gas-to-dust ratio was fixed to be 1:100. 

The best fitting model disc has a mass of 0.01\,M$_\odot$, which agrees with 
the literature data \cite{Natta:discmass} within the uncertainties, and an 
outer radius  of 320\,AU. The rim temperature was 1800\,K which is equivalent 
to $\sim 0.6$\,AU inner radius for the assumed stellar properties. The surface 
density was $\Sigma(r)\propto r^{-0.9}$. The inclination of the disc was set
to about 70$^{\circ}$, which is close to an edge-on configuration, 
as proposed by many authors for the UXORs (Dullemond et al. 2003). In order to match 
the observed optical colours, an interstellar redenning of $A_V=0.74^m$ had also to be 
adopted. The resulting fit to the spectral energy distribution is displayed in
Fig.\,\ref{fig:modsed}.

\begin{figure}
\centering
\includegraphics[width=88.1mm]{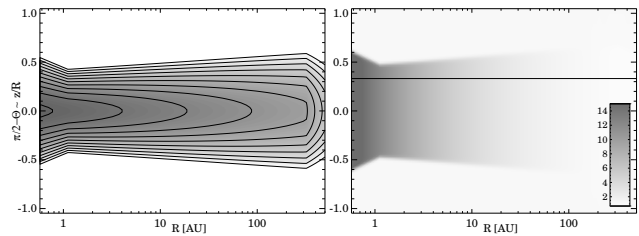}
\caption{Density structure of the model disc. {\it Left:} 
density distribution of the best fitting stationary model of the SED of
May 31, 1997. The density contours are plotted as a function of the radial
distance and the polar angle. The puffed-up inner rim ($r\leq 1.5$\,AU)
is opaque below a polar angle of $\sim$15$^{\circ}$, puting most of the disc
into the shadow. The line-of-sight is at about 20$^{\circ}$. {\it Right:}
Change of density in the disc between the two extreme models 
with surface density exponents -0.9 and -0.4. The solid line marks the
line-of-sight.}
\label{fig:vari_struct}
\end{figure}

The resulting geometry is a disc whose vertical pressure scale height
increases with the radial distance (see Fig.\,\ref{fig:vari_struct}, 
{\it left panel}). 
The height of the disc photosphere (defined as $\tau=$1 for stellar radiation at
$\lambda=$ 0.55 $\mu$m) also increases, approximately as $h/r \propto r^{1/10}$,
marking a moderately flared geometry. 
The disc has an inner hole at $r<0.6$\,AU, surrounded by a puffed-up inner 
rim created by the direct illumination of the inner edge of the disc by the star. 
The disc of SV\,Cep is basically self-shadowed, however, due to the flaring, 
some material -- especially at larger radii --
is out of the shadow, and can be illuminated directly
by the star (Fig.\,\ref{fig:vari_struct}). These cold outer regions, far above
the disc mid-plane, are optically thin for the stellar radiation, otherwise
the star would not be visible from the inclination angle of 70$^{\circ}$.

Our model fits the continuum radiation over 3 orders of magnitudes in wavelengths,
from the optical to the far-infrared, reasonably well. This result verifies that the 
assumed self-shadowed geometry is a very good description of the circumstellar
environment. Local differences, however, appear in the 10--20\,$\mu$m range where
the model cannot reproduce the observed silicate emission features. The reason is
that in the  model the surface layer of the disc immediately behind the rim, 
where the silicate emission arises, is not illuminated. In reality,   this
region of the disc might be irradiated indirectly by stellar  photons scattered
in the rim photosphere, producing the observed emission features. An other
possible explanation for the silicate emission features could be an  optically
thin halo around the star\,+\,disc system (see below). Also, our disc model is not 
able to fit the observed 1.3\,mm flux. The age of SV Cep (4\,Myrs) 
suggests that significant fraction of big grains ($r\sim 1\,mm$) may be 
settled in the disc midplane, which radiate at this wavelengths. Since
these particles are in the disc midplane and have more or less grey opacities,
their radiation should not vary.\\

\noindent{\bf Modelling light variations.}
After finding a stationary model which fits the SED of SV\,Cep at a  reference
epoch, one may try to model the observed flux variations. A possible origin
of the flux variations can be that the illumination source of the circumstellar disc,
e.i. the star itself, changes. However theoretical models predict the presence 
of an instability strip on the Hertzsprung-Russel Diagram (HRD) \cite{Boehm}, the possible
periods are in the order of an hour or even less. This periods are too short
comparing with the observed timescale ($\sim$years) in the light curves of SV Cep.
Thus our working hypothesis
is that the optical variations are caused by long-term eclipses in the inner
rim, produced by some effects in the rim which change its height along a
significant  fraction of its perimeter (rather than along the line of sight 
only). The eclipse nature of the phenomenon is supported by observations 
of Rostopchina et al. \cite{Rostopchina}, who showed that SV\,Cep turns  
redder when becoming fainter. This hypothesis predicts some consequences 
in the infrared regime. A higher inner rim, while obscuring the star, 
absorbs more stellar radiation which is re-emitted  then as 2--8$\mu$m 
thermal emission.  At the same time a higher rim results in stronger 
shadowing of the outer regions. Since the disc around SV\,Cep is thought 
to be passive, a stronger shadow results in a decreased radiation from the 
outer parts of the disc in the far-infrared domain. 
Because both the inner rim and the optically thin outer part of the disc
are expected to react on the changes of the stellar illumination almost
instantaneously, both the optical vs. mid-IR
anti-correlation and the optical vs. far-IR correlation should be observable.

The prediction that the optical and far-infrared emission changes in the same
direction is fully consistent with our observational results which exhibit
correlation between the optical and far-infrared fluxes 
(Fig.\,\ref{fig:mod_corr}). The predicted optical vs. mid-infrared
anti-correlation seems also be consistent with the data, though in this
case the observational results are less conclusive.

We tried to model our changing rim scenario quantitatively with the RADICAL and
the RADMC.
Since these codes are time-independent, stationary codes, our concept was to
model the variability through a sequence of quasi-stationary states. In this
process the variable height of the rim was simulated by changing the amount of
matter in the rim area (i.e. the surface density at the inner edge of the
disc). Here we assumed circular symmetry, i.e. the density of the rim was
increased or decreased along its complete perimeter simultaneously. The
vertical Gaussian density profile of the disc has a pressure scale height 
which is determined by the temperature and independent of the density 
\cite{Dullemond:hole}. Thus a higher
surface density with the same pressure scale height causes higher volume
density and higher opacity in the rim, and the photosphere of the rim moves up
in vertical direction. In the course of modelling the variation of surface
density  at the inner edge of the disc was realized by changing the exponent of
the power law of the surface density between -0.9 and -0.4. 
Figure\,\ref{fig:vari_struct} ({\it right panel}) shows the change of density
in the whole disc between the two extreme models  with exponents -0.9 and -0.4.
The plot demonstrates that a modification of the exponent does not alter the
structure of the outer disc considerably (the surface density was constant
within 10\%), but could efficiently be used for fine-tuning  the surface
density  in the rim. 

By gradually changing the exponent of the surface density (we emphasize that
it was the only parameter tuned) we created a sequence of spectral energy
distributions, from which we produced synthetic light curves and correlation 
diagrams. The results revealed that the amplitude of the infrared variations 
does not depend on  the viewing angle, but the amplitude of the optical changes 
does. In order to match the relative amplitude of the optical and far-IR 
variations we played with the inclination angle; the best fitting value was
$i = 71^{\circ}$. The dashed lines in Fig.\,\ref{fig:mod_corr} mark linear fits
to the synthetic correlation diagrams. The plots demonstrate that our numerical 
simulation of the  changing rim scenario is able to reproduce the observed trends 
of the light variation in the optical, near- and far-infrared domains simultaneously. 

\begin{figure}
\centering
\includegraphics[width=88.1mm]{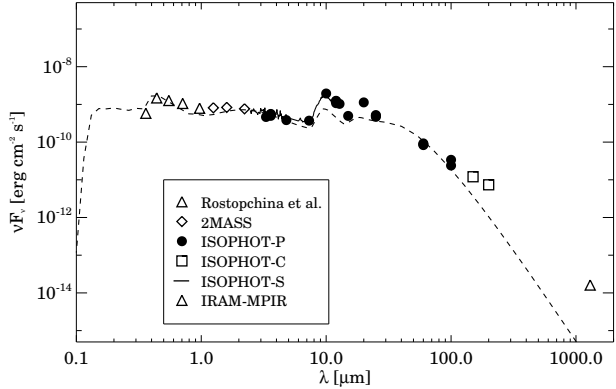}
\caption{SED of SV Cep. The measured data points are corrected for an
interstellar reddening of 0.57$^m$. Overplotted is the best fitted disc$+$envelope  model.}
\label{fig:envsed}
\end{figure}

However, the model is not able to explain the mid-infrared  behaviour of 
SV\,Cep. The predicted flux variations at 12 and 25\,$\mu$m would have similar 
amplitude and direction (i.e. correlation with the optical) than was seen at 
100\,$\mu$m, in contrary to the observed almost constant brightness at these 
wavelengths. It is interesting to note that this is the same wavelength range 
in which the model SED was not able to fit the observed emission features.

It is a question whether the hypothetic changes in the amount of matter in the 
inner rim are physically reasonable?
The rim contains less than 1\% of the total disc mass. In our simulations we had 
to vary the rim mass between $10^{-4} - 10^{-5} M_{\odot}$ on a time-scale of a few 
years. If this change was explained by varying accretion, the accretion rate would 
be in the range of $10^{-5} M_{\odot}$/yr; an unexpectedly high value for an old 
UX\,Orionis star. An alternative explanation would be to assume that the
pressure scale height of the rim may also change due to some dynamic 
instabilities which affect a significant portion of the rim at the same time.
Further analyses are needed to clarify the possible nature of the 
variations of the rim structure.

\subsubsection{Disc$+$envelope model}

\begin{figure}
\centering
\includegraphics[width=88.1mm]{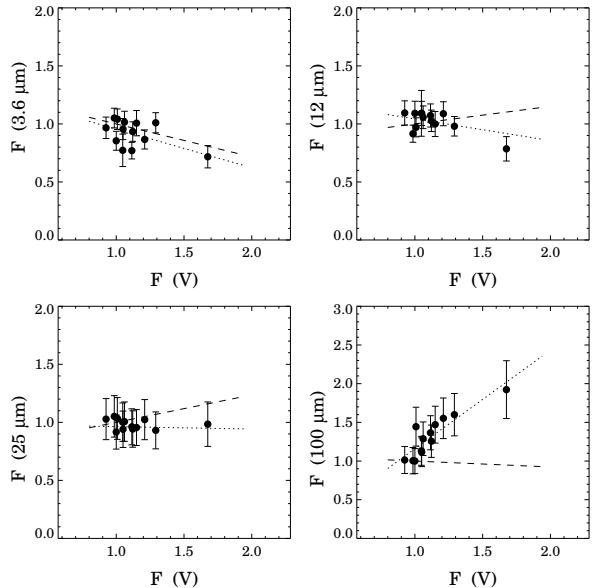}
\caption{Correlation diagrams between the optical and 
infrared light variations. All light curves are normalized
for the reference day of May 31, 1997. Dotted lines are linear fits to the 
observations; dashed lines show the same for synthetic light curves 
obtained from modelling using a disc$+$envelope system (Sect.\,\ref{modelling}).} 
\label{fig:env_corr}
\end{figure}

Since the pure disc model is not able explain the observed behaviour of SV Cep at mid-infrared
wavelengths (12--25\,$\mu$m) we tried to find an alternative solution. The lack of variability 
suggests that the heating source of the matter emitting at these wavelengths is 
independent of the height of the inner rim. An obvious way to meet this criterion is 
the introduction of an optically thin envelope around the system. 
Since the envelope is a spherical structure it can be illuminated by the central star 
continuously, the inner rim has no significant effect on the irradiation of the envelope.
Moreover an optically thin envelope could produce the observed huge silicate emission
features around 10 and 20\,$\mu$m.\\

\noindent{\bf Modelling the SED.}
On purely physical grounds we are able to constrain some property of the envelope.
If the inner radius would be smaller than 10\,AU, the variability would disappear 
around 3.6\,$\mu$m, due to the contribution of the envelope emission to the SED. 
On the other hand, if the inner radius would be greater than 30--40\,AU, the dust 
temperature would not be high enough to produce the emission feature at 10\,$\mu$m. 
Thus we set the inner radius to 15\,AU, while the outer radius 
of the envelope was set to 500\,AU which was also the outer radius of the grid. 

There are disc$+$envelope models in the literature which explain the 
2--8\,$\mu$m emission in Herbig Ae/Be stars by invoking a compact ($\sim$10\,AU), 
tenuous (${\tau}_{V}{<}{\sim}0.4$) dusty halo around the disc inner regions, rather 
than by assuming an inner rim (Vinkovic et al.\,2006). Since the 
observations showed variability at the 2--8\,$\mu$m region, which cannot be 
explained in the case of an envelope, we do not adopt the model of Vinkovic et al.\,\cite{Vinkovic}.

Dust grains of
two different sizes (0.01 and 0.1\,$\mu$m) were used to produce the emission features around 10 and 
20\,$\mu$m. The volume density distribution used in the model ($\rho(r)\propto r^{-2}$) 
was typical of an infalling envelope. The only fitted parameter of the envelope was
the volume density at the outer radius of the envelope; its best fit value was 
$\rho_0=3.0\times10^{-20}$\,g/cm$^3$. 
The applied disc parameters were the same than in the case of the pure disc model 
except for a few. The best fit disc model had a mass of $10^{-3}$\,M$_{\odot}$ and the grain sizes
were somewhat larger (0.75 and 3\,$\mu$m) than in the envelope. The lower disc mass results
in a geometrically thinner disc, thus the inclination angle in the best fit model was modified to 
76$^{\circ}$ in order to obtain the same optical variability than in the pure disc model. 

The resulting fit to the spectral energy distribution is displayed in 
Fig.\,\ref{fig:envsed}. It can be seen that the fit of this model to 
the continuum part of the SED is as good as in the case of the pure disc model.
In the mid-infrared domain around 10\,$\mu$m the fit of the disc$+$envelope
model is remarkably better than that of the pure disc model.\\

\noindent{\bf Modelling light variations.}
The disc$+$envelope model can successfully reproduce the observed SED, but we also 
have to investigate the predicted light variations of this model at different wavelengths. 
The optical variability was simulated in the same way as in the case of the 
pure disc model (i.e. varying of the height of the inner rim). During this simulation 
the power index of the surface density distribution was tuned between -0.9 and -0.4.
Then correlation diagrams were constructed from the synthetic light curves of  
different wavelengths. Figure \ref{fig:env_corr} shows that the disc$+$envelope model
is able to reproduce quantitatively the observed flux variations both in the optical and the 
2--8\,$\mu$m region and the anti-correlation between them.
The model predictions in the mid-infrared domain around the silicate features are much
closer to the observations than in the case of the pure disc model. 
Since the envelope radiates most of its energy around 10 and 
20\,$\mu$m and its emission is practically constant, the lack of flux variations
at these wavelengths is not surprising. 
However in the far-infrared domain this model seems to be in clear conflict with the
observations: it predicts an almost completely constant
flux in contrary to the observed prominent variability at these wavelengths. 

The reason of the lack of variability at far-infrared wavelengths is probably the heating of the outer
parts of the disc by the envelope. In a disc$+$envelope model the outer regions of the circumstellar
disc have two heating sources: radiation from the central star and from the envelope. The shadowing
effect of the inner rim in this model is so strong, that the outer parts of the disc absorb
more energy from the envelope than from the central star. Further than 
10\,AU from the star this process results in higher temperatures in the disc compared to a pure
disc model. Since the flux emitted by the envelope is constant in time, the far-infrared flux 
will also be constant. 

\subsection{Comparison of the models}

A systematic comparison of fitting the observed phenomena in the pure disc and the 
disc$+$envelope models is presented in Tab.\,\ref{tab:summary}. It can be seen that neither
of the models is able to explain all the phenomena together. As was mentioned in 
Sec.\,\ref{sec:mod_sed_disc} the quality of the fits could probably be improved by taking
into account the scattering of stellar light which provides extra heating to the disc 
region where the silicate emission features originate. Modelling the far-infrared variability
is, however, still a critical issue. In the disc$+$envelope model
the temperature distribution of the outer parts of a self-shadowed disc is determined by the 
absorbed energy from the envelope. As discussed before, radiation of the envelope is not 
expected to show temporal variation, thus it is hard to find a way to produce far-infrared 
variability in such a model. Since far-infrared variability was definitely detected
in our dataset, this is a strong argument against the disc$+$envelope model. The pure disc
model might be a better starting point to model successfully the structure and behaviour 
of the circumstellar environment of SV Cep.

\begin{table}
\begin{center}
\begin{tabular}[width=88.1mm]{lcc}
\hline
Observed phenomenon & Pure disc & Disc$+$envelope\\
\hline
SED continuum & Yes & Yes \\
SED emission features & No & Yes \\
Optical variability & Yes & Yes \\
NIR variability & Yes & Yes \\
MIR variability & No & Yes \\
FIR variability & Yes & No \\
\hline
\end{tabular}
\caption{Summary of the relations between the observed phenomena and the predictions of the two
different models.}
\label{tab:summary}
\end{center}
\end{table}

\section{Summary and Outlook}

We investigated the long-term optical-infrared variability of SV\,Cep, and
tried to explain it in the context of the self-shadowed disc model proposed
for UX\,Orionis stars.
A 25-month monitoring programme, performed with the Infrared Space
Observatory in the 3.3--100\,$\mu$m wavelength range, was carefully analysed,
and the infrared light curves were correlated with the behaviour of
SV\,Cep in the V-band.
A remarkable correlation was found between the optical and the far-infrared
light curves. In the $2-8\,\mu$m regime the amplitude of variability is lower,
with a hint for a weak anti-correlation with the optical changes. No observable
flux changes were recorded around 25\,$\mu$m.

In order to interpret
the observations, we modelled the spectral energy distribution of 
SV\,Cep assuming a self-shadowed disc with a puffed-up inner rim, using a
2D radiative transfer code. The simulations revealed that the main
body of the SV\,Cep disc is located in the shadow of the inner rim (confirming
the expected self-shadowed geometry), but a moderate flaring is also present in
the outer disc which makes it possible that some cold tenuous regions far above
the disc midplane can be directly illuminated by the star. 
We found that by modifying the height of the
inner rim, the wavelength-dependence of the long-term optical-infrared variations
is well reproduced, except the mid-infrared domain. The origin of variation of the 
rim height might be temporal changes in the accretion rate, but other mechanisms 
which alter the pressure scale height of the rim might be necessary to invoke.
In order to model the mid-infrared behaviour we tested adding an optically thin 
envelope to the system, but this model failed to explain the far-infrared variability.

We demonstrated that infrared variability is a powerful tool to 
discriminate between existing models. 
The proposed mechanism of variable rim height may not be restricted to
UX\,Orionis stars, but might be a general characteristic of
intermediate-mass young stars. However, in star with pole-on view no
optical changes are expected. Further infrared variability studies
among Herbig Ae/Be stars would be needed to confirm this prediction.

\section*{Acknowledgements}
The authors thank the referee, B. Whitney for her comments. 
This paper was partly based on observations with ISO, an ESA project with
instruments funded by ESA member states (especially the PI countries France,
Germany, the Netherlands and the United Kingdom) with participation of ISAS and
NASA. The ISOPHOT data presented were reduced using the ISOPHOT Interactive
Analysis package PIA, which is a joint development by the ESA Astrophysics
Division and the ISOPHOT Consortium, lead by the Max-Planck-Institut f\"ur
Astronomie (MPIA). 
The work was partly supported by the grant OTKA K\,62304 of
the Hungarian Scientific Research Fund. 



\label{lastpage}

\end{document}